\documentclass[12pt,preprint]{aastex}
\shorttitle{A Transit Detection Method}
\shortauthors{Weldrake $\&$ Sackett}
\begin{document}
\title{A Method for the Detection of Planetary Transits in Large Time-Series Datasets}

\author{David T F Weldrake} 
\affil{Research School of Astronomy and Astrophysics, Mount Stromlo Observatory, Cotter Road, Weston Creek, ACT 2611, Australia}
\email{dtfw@mso.anu.edu.au}

\and

\author{Penny D Sackett}
\affil{Research School of Astronomy and Astrophysics, Mount Stromlo Observatory, Cotter Road, Weston Creek, ACT 2611, Australia}
\email{psackett@mso.anu.edu.au}

\begin{abstract}
We present a fast, efficient and easy to apply computational method for the detection of planetary transits in large photometric datasets. The code was specifically produced to analyse an ensemble of 21,950 stars in the globular cluster 47 Tucanae for the photometric signature indicative of a transiting Hot Jupiter planet, the results of which are the subject of a separate paper. Using cross correlation techniques and Monte Carlo tested detection criteria, each photometric time-series is compared with a database of transit models of appropriate depth and duration. The algorithm recovers transit signatures with high efficiency while maintaining a low false detection probability, even in rather noisy data. The code has been optimized and with a 3GHz machine is capable of analysing and producing candidate transits for 10,000 stars in 24 hours. 

We illustrate our algorithm by describing its application to our large 47 Tuc dataset, for which the algorithm produced a weighted mean transit recoverabilty spanning 85$\%$ to 25$\%$ for orbital periods of 1$-$16 days (half the temporal span of the dataset), despite gaps in the time series caused by weather and observing duty cycle. The code is easily adaptable and is currently designed to accept time-series data produced using Difference Imaging Analysis.

\end{abstract}
\keywords{stars:planetary systems - occultations - methods:data analysis}

\section{Introduction}
The periodic transits of extrasolar planets as they pass across the face of their parent star presents an important diagnostic tool for planetary detection and characterisation. The method allows a direct measurement to be made of key system parameters, including the orbital inclination and the orbital period, and provides the only way at present to directly measure the planetary radius. Radius measurements constrain planetary migration history and evolution, and also allow comparison of observations to models of planetary atmosphere and composition \citep{Burrows2000,Bar2003,Burrows2004}. When coupled with radial velocity measurements, an accurate planetary mass and density can be derived \citep{Char2000,Brown2001}. Furthermore, the transit method can detect planets in regions outside the immediate solar neighbourhood and thus can be used to search for planets in distant locations of high stellar density, including globular clusters \citep{Gil2000,Weld2005}, open clusters \citep{Moch2002,Moch2004} and dense Milky Way starfields \citep{Udal2002,Udal2003}. 

The use of transits for planetary detection was first suggested by \citet{Struve52} and studied in detail by \citet{BR1984}, but due to technological limitations was not implemented until \citet{SD1995} and \citet{Doyle1996}, who used the method to search for planets in orbit around the M-dwarf binary CM Draconis. \citet{Deeg2000} announced the detection of a variation in eclipse minimum times in this system, which could indicate the presence of a third body of planetary mass. Further observations are required for confirmation; as yet, no transits in CM Draconis have been observed. 

Interest in planet transits received a considerable boost after the photometric detection of HD209458b \citep{Char2000,Henry2000}, a short-period Jupiter-mass planet in orbit around a solar-type star, spurring many other transit searches in recent years \citep{Horne2003}. Transits have a certain geometrical probability (P$_{geo}$) of occuring for any given star, depending on the stellar radius R$_{\ast}$ and the orbital semi-major axis $\it{(a)}$ via: 
\begin{center}
\begin{equation}
P_{geo} = \frac{R_{\ast}+R_{p}}{a} \sim \big(\frac{R_{\ast}}{a}\big)
\end{equation}
\end{center}
\noindent where R$_{p}$ is the planetary radius. Blind transit searches are most sensitive to large planets with very small orbital radii, the so-called Hot Jupiter planets. Typical Hot Jupiter transit depths are $\sim$0.015 mag \citep{Char2000}, with a duration of a couple of hours, and thus are challenging to detect with data taken in average ground-based conditions, for which the transit depth would be comparable to the photometric noise. 

 As a general rule, transit searches involve dedicated small telescopes that monitor wide fields of view, with expected transit yields of several per year (see e.g. \citet{Hidas2003}. An all-sky transit survey has recently been described by \citet{Deeg2004}, which could lead to the detection of many new planets in the near future. A particularly successful search has been undertaken by the OGLEIII group \citep{Udal2002,Udal2003}, who have already identified 137 transit candidates. The nature of the transit signatures is currently being determined via vigorous radial velocity follow-up campaigns; of these OGLEIII candidates, four objects have been confirmed as planets \citep{Kon2003a,Bouchy04,Kon2004,Pont2004}. Further to these, \citet{Alonso2004} announced the identification of TrES-1, the first planet transit to be identified with the Trans-Atlantic Exoplanet Survey. Due to the precise alignment (viewing geometry) required, a large ensemble of stars must be sampled in order to detect the fraction of any planets that will transit their host as seen from Earth. With such large datasets, in which many thousands of stars must be simultaneously sampled, it is necessary to employ a computational detection algorithm that can efficiently identify candidate detections. 

We present here a detection algorithm developed and tested to search a dataset of 21,950 stars in the globular cluster 47 Tucanae for transiting planets, the results of which are presented in a companion paper \citep{Weld2005}. The code was tested extensively to maximize the detection of modeled transits of appropriate depth and duration, while keeping the false detection rate to an acceptable level. The result is an effective, fast method of transit detection that can be easily modified.

Section 2 presents an overview of our detection algorithm, along with details of the model creation and detection statistic. Section 3 explains the detection criteria that separate real candidates from transit-like systematic effects in the presence of noise. Section 4 describes the Monte Carlo simulations we have performed to test the transit recoverability rates in our dataset and determine the false detection rates. Section 5 presents the summary and conclusions.

\section{General Method for Transit Detection}
In order to search for periodic transit signatures in our 47 Tuc dataset, we employed a variation of the Matched Filter Algorithm (MFA). This method of signal detection involves comparing the data to a series of appropriate models, and was first suggested for use in transit searches by \citet{Jenk96}. The Matched Filter approach has been described in the literature as the best method for use when specifically applied to search for transits \citep{Ting03a,Ting03b,Kov2002} and has been used for several transit searches, for example, \citet{Gil2000} and \citet{Bruntt03}. The method, as we have implemented it, assumes a simple square-well transit shape, which is a valid assumption when searching for a signal very close to the noise level, for which the shape of the short ingress/egress phase of the transit is essentially unresolved. The algorithm searches for multiple transits spanning a large range of periods and transit start-times within a parameter space determined by characteristics of the dataset. With a total observing window of $\it{L_{w}}$ days, the theoretical upper limit of orbital period to contain three detectable (and hence confirmed) transits is $\it{L_{w}/2}$ days; we take this as the upper limit of the orbital period spanned in the search. 

Superior results will be obtained if lightcurves containing obvious systematic effects are removed or the systematic variations reduced before they are submitted to the search algorithm. The lightcurves used in our 47 Tuc dataset were produced via difference imaging analysis, originally described as an optimum point-spread function (PSF) algorithm by \citet{AL98} and later modified by \citet{Woz2000}. Difference imaging analysis automatically removes many of the unwanted systematic effects caused by PSF variations over the observational time span, and is particularly useful in crowded stellar fields. Other methods to reduce the contamination caused by systematic effects are contained within our algorithm itself and are described in Section 3. Throughout, we will use results from our 47 Tuc dataset \citep{Weld2004}, hereafter WSBFa, to illustrate the detection algorithm.

\subsection{Model Creation and Detection Statistic}
We begin by producing a large number ($\it{N_{mod}}$) of model transit lightcurves with which to compare the data. Each model has a transit period of $\it{P_{mod}}$ days, a transit depth of $\it{D_{mod}}$ magnitudes, and a transit duration of $\it{d_{mod}}$ hours. These values are chosen within a range of Hot Jupiter transit depths and durations expected to be detectable in the dataset. Further to these parameters, each model is characterised by $\it{\tau_{shift}}$, the time from the first data point to the beginning of the first transit.

As an example, Fig.\ref{3.15dnogaps} shows a transit model with $\it{P_{mod}} =$ 3.15d, $\it{D_{mod}} =$ 0.02 mag, and $\it{d_{mod}} =$ 2 hours covering the whole of a $\it{L_{w}}$$ =$ 30.4 day observing window. This $\it{P_{mod}}$ is typical of the most common Hot Jupiter planet orbital period from the sample currently known in the Solar Neighbourhood\footnote{From the Extrasolar Planet Encyclopedia http://www.obspm.fr/encycl/encycl.html}, and is used as an example to illustrate the visibility of such a transit in a perfect dataset. In this illustration there are no gaps in temporal coverage, which would be obtained only in perfect weather conditions experiencing no diurnal effects. 

The total number of models, $\it{N_{mod}}$, is the product of the total number of period steps, $\it{L_{w}/2\Delta P_{mod}}$, where $\it{\Delta P_{mod}} = d_{mod}/2$ and the total number of transit start time steps, $\it{P_{mod}/\tau_{shift}}$. The net effect of having these $\it{\Delta \tau_{shift}}$ steps is to slide each model (with a particular $P_{mod}$) across the data. The statistic we use to compare the model lightcurves to the data and test for the presence of transits is a cross-correlation function, $\it{C(P_{mod},\tau_{shift})}$, of the form:
\begin{center}
\begin{equation}
C(P_{mod},\tau_{shift}) = \sum_{i=1}^n D(t_i) M(P_{mod},\tau_{shift}+t_i) \Delta t
\end{equation}
\end{center}
\noindent where the obtained lightcurve consists of $\it{i}= 1 \rightarrow \it{n}$ points; $C(P_{mod},\tau_{shift})$ is the value of the cross correlation function for a transit model with period $P_{mod}$ beginning at time $\tau_{shift}$, $D(t_i)$ is the deviation (in $\Delta$ magnitudes) of the data from the mean value of the lightcurve at the observational time $t_{i}$, $M(P_{mod},\tau_{shift}+t_i)$ is the deviation (in $\Delta$ magnitudes) of the model at this same time, and $\Delta$t is taken to be $\Delta \tau_{shift}$. The output of this detection statistic is therefore a total series $\it{N_{mod}}$ values of $\it{C(P_{mod},\tau_{shift})}$ per lightcurve. Note that by taking $\it{\Delta t = \Delta \tau_{shift}}$, we are weighting every data point equally in the cross correlation. A discussion on the inclusion of non-equal weighting shall be made in section 3.

The time series data ($\it{i}=1 \rightarrow \it{n}$) are first shifted so that their mean equals zero, allowing a direct comparison between the data and each transit model (which has out-of-transit points set to zero). Note that only data points lying inside the model transit are used for the analysis, as, by definition, points outside transit are zero in the model $M(P_{mod},\tau_{shift}+t_i) \equiv 0$ and thus do not contribute to $C(P_{mod},\tau_{shift})$, saving computational time. Models with short periods have more points inside transit and hence have higher significance detections than those with longer periods. If a simulated transit is added to the actual lightcurve data, a $\it{C(P_{mod},\tau_{shift})}$ will be a miximum for the model that best represents the transit.
 
In summary, each lightcurve is compared to a total of $\it{N_{mod}}$ transit models, spanning a period $\it{(P_{mod})}$ range of 1 $\rightarrow \it{L_{w}}/2$ days, with $\it{\Delta P_{mod}}$ steps of $\it{d_{mod}/2}$, and a $\tau_{shift}$ range from 1$\rightarrow P_{mod}$. The step size $\Delta \tau_{shift}$ should be chosen with regard to the temporal sampling particular dataset, and is best chosen to be 1/2 the transit duration or smaller. 

This total database of models and their associated $C(P_{mod},\tau_{shift})$ values for each lightcurve allow us to completely search an individual lightcurve for any transit-like feature that could theoretically be contained within it, while optimising the time needed for the code to run. By calculating the root-mean-square scatter, $\it{\sigma_{N_{mod}}}$, of the output $C(P_{mod},\tau_{shift})$ points over all models, a measure can be made of the significance $S(P_{mod},\tau_{shift})$ for a given lightcurve. Namely,
\begin{center}
\begin{equation}
S(P_{mod},\tau_{shift}) \equiv \frac{C(P_{mod},\tau_{shift})}{\sigma_{N_{mod}}}
\end{equation}
\end{center}
\noindent where $\it{C(P_{mod},\tau_{shift})}$ is the value of the cross-correlation function of a given model and $\sigma_{N_{mod}}$ is given by:
\begin{center}
\begin{equation}
\sigma_{N_{mod}}= \sqrt {\frac {\sum_{P_{mod}}\sum_{\tau_{shift}} \bigg[ C(P_{mod},\tau_{shift}) - \overline{C(P_{mod},\tau_{shift})} \bigg]^2}  {{N_{mod}}}}
\end{equation}
\end{center}
\noindent where $\overline{C(P_{mod},\tau_{shift})}$ is the mean cross-correlation value over all models for a given lightcurve. Put more simply, this measure converts the raw $\it{C(P_{mod},\tau_{shift})}$ values from their original units into multiples of the total rms scatter of the $\it{C(P_{mod},\tau_{shift})}$ distribution, providing an output that is easier to interpret.

\subsection{Application to the 47 Tuc Dataset}
For the 47 Tuc dataset (WSBFa), we adjusted the values of the code parameters discussed earlier for speed and maximum recovery. First, our total observing window $\it{L_{w}}$ is 30.4 days, giving an upper limit to the Hot Jupiter orbital period of 15.2 days for secure detection. Fig.\ref{3.15dgaps} shows the actual 47 Tuc dataset temporal coverage with the same transit model as in Fig.\ref{3.15dnogaps}, and highlights the serious adverse effect of diurnal observing on transit visibility. To analyse the 47 Tuc dataset, we used 1501 $\it{\Delta P_{mod}}$ steps, each of 0.01 days increment, and 755 $\it{\Delta \tau_{shift}}$ increments, each of 0.04 days, yielding a total $\it{N_{mod}}$ of 1,133,255 multiple transit models to compare to each time series. 

Fig.\ref{3.15ddata} shows a synthetic transit (with the same $\it{P_{mod}}$, $\it{D_{mod}}$ and $\it{d_{mod}}$ as the model presented in Figs.\ref{3.15dnogaps} and \ref{3.15dgaps}) added to an actual randomly-selected lightcurve from the 47 Tuc dataset with 0.015 magnitude photometric uncertainty. The transit is very easily seen in the data. When phase-wrapped at the peak detected period (3.15d, the actual true period), the result is Fig.\ref{3.15dpwrap}. The transit is clearly seen at phase $\Phi=$ 0 and 1. 

An example of a transit added to a 47 Tuc star of poor photometric quality (0.035 mag rms) is shown in Fig.\ref{3.15ddata_0.03} and is phase-wrapped in Fig.\ref{3.15dpwrap_0.03}. The model has the same $\it{P_{mod}}$ and $\it{d_{mod}}$ but a transit depth $\it{D_{mod}}$ of 0.03 magnitude. The code detected this transit successfully, with a high significance, which is impressive considering the noise level of the data. 

Fig.\ref{ccfplot} displays the cross correlation output for the models presented in Figs.\ref{3.15ddata} and \ref{3.15ddata_0.03} in order to better illustrate the detection method. The detection significance, $\it{S(P_{mod},\tau_{shift})}$, is indicated for both modeled transits for the true transit period, $\it{P_{tran}}$, and twice the true period, 2$\it{P_{tran}}$. The points show the output significance of these two models with a correct $\it{P_{tran}}$ and correct $\it{\tau_{shift}}$ (CPCT), correct $\it{P_{tran}}$ with incorrect $\it{\tau_{shift}}$ (CPIT) and incorrect $\it{P_{tran}}$ with correct $\it{\tau_{shift}}$ (IPCT). It is clear that the correct model has the highest significance, and the transit can be seen visually when phase-wrapped to this period. 

Fig.\ref{mfaoutput} shows the significance values of models tested against a  $\it{P_{tran}}=$1.149d transit with a depth of 2$\%$ (0.02 mag) added to a (random) actual 0.015 magnitude uncertainty lightcurve from our 47 Tuc data. An extremely strong detection with $S(P_{mod},\tau_{shift})=22$ is found at the true $\it{P_{tran}}$ although aliasing is apparent at integer multiples of this period. The bottom half of the figure shows the ouput for the same lightcurve but with no transit added. Only three data points are found above the detection criterion $\it{S(P_{mod},\tau_{shift})\ge S_{cr}}$ that we have placed and describe in the next section. These three points are caused by random scatter in that lightcurve.

\section{Detection Criteria in the Presence of Noise}
In order to separate real detections from transit-like false detections caused by variations inherent to the data, detection criteria are required. Both statistical and systematic variations can produce high significance $\it{S(P_{mod},\tau_{shift})}$ that do not correspond to actual real transits. (See, for example Fig.\ref{mfaoutput}). Consequently, the first detection criterion is a lower limit on $\it{S(P_{mod},\tau_{shift})}$; the higher $S(P_{mod},\tau_{shift})$, the better is the correlation between the model and the data. However, this single criterion is not sufficient, as many lightcurves have systematic effects that lead to spuriously high $\it{S(P_{mod},\tau_{shift})}$ values, producing points that lie in the detection regime. A further criterion is required to reduce these spurious detections. 

Due to aliasing effects a real detection of multiple transits will yield a larger number of different models with $S(P_{mod},\tau_{shift})$ that satisfy the first criterion. This has been verified by testing the dataset. Therefore the second criterion involves measuring the actual number of models, $N_{p}$, that lie above the first $S(P_{mod},\tau_{shift})$ detection criterion.

As well as the $S(P_{mod},\tau_{shift})$ and $N_{p}$ criteria, false detection rates can be reduced by considering the uncertainty of the photometry. A very small number of strongly discrepant points in the time-series with small formal errors can account for a large portion of false detections. We have found that the majority of these spurious points are caused by bad columns and other CCD blemishes, or by suddenly differing observing conditions. As long as the deviation of these discrepant points is larger than would be expected for a transit, we reset the $\it{D(t_i)}$ value of that point to the mean of the data. The consequences of this shall be discussed further when considering the application of these criteria to the 47 Tuc dataset.

The detection process is complicated by the varying observational conditions typical of long time series of photometric data. These can produce pseudo periodic signals with an associated increase in photometric measurement errors. In order to reduce these effects, the contribution of each datapoint before it was used in the $C(P_{mod},\tau_{shift})$ calculation was weighted by the size of its photometric uncertainty, with the standard weighting scheme:

\begin{center}
\begin{equation}
W_{i} = \frac{1}{\sigma_i^2},
\end{equation}
\end{center}
\noindent where $W_{i}$ is the point weight and $\it{\sigma_i}$ is the errorbar associated with the i$^{th}$ point. The result is that points with large errorbars are given a small weight and hence do not add significantly to the final $\it{C(P_{mod},\tau_{shift})}$ for that model. By incorporating both of the $S(P_{mod},\tau_{shift})$ and $N_{p}$ detection criteria, and incorporating the outlier removal and $\it{W_{i}}$ weighting scheme, the real detections and false detections in the time-series were kept to acceptable levels.

\subsection{Detection Criteria for the 47 Tuc Dataset}
We now discuss the application of these criteria to the 47 Tuc dataset, and illustrate their effect on the final transit candidate lists. The data were split into two bins for separate analysis, distinguishing lightcurves with relatively low photometric scatter ($\le$0.02 mag), and those with somewhat higher scatter (0.02$\le$rms$\le$0.04 mag). We found that applying slightly different values of the detection criteria to the two bins we could keep the detection levels high and false detection levels low.

We performed Monte Carlo tests, adding model transits of various depths and durations to actual dataset lightcurves with differing photometric uncertainties to determine the maximum photometric scatter for which the algorithm could reasonably detect transits with depths as large as $\it{D_{tran}} =$ 0.03 mag. The expected depth of a transit is dependent upon stellar magnitude and this value is the expected transit depth for stars at the lower limit of our search range, as described in \citet{Weld2005}. Stars with scatter greater than 0.04 mag were found to suffer from large false detection rates and an unacceptably low transit recoverability rate. With this lower limit, the total number of lightcurves with rms $\le$0.04 mag in the 47 Tuc dataset is 21,950, allowing a statistically robust sample for analysis. 

For the better quality lightcurves, only those models with $S(P_{mod},\tau_{shift}) \ge S_{cr}=$6.5 (marked by a dotted line in Fig.\ref{mfaoutput}) are passed onto the second detection criteria. For poor quality stars in the second bin, this criterion is $S_{cr} \ge$ 7.0. The second criterion $\it{N_{p}}$ was then calculated, and only those lightcurves with $\it{N_{p}}$$>$60 and $>$50 points, for the high and low quality bins, respectively, were considered further. These specific numbers for the detection criteria were determined via a trial and error process, where the numbers were altered and the subsequent results noted. The final choice was made by maximizing the transit detection while keeping false detection rates to a low level.

We chose to remove outliers outside 3.5 times the rms lightcurve scatter. This could remove real transits with depths $\it{D_{tran}}>$ 0.035 mag in our best quality main sequence lightcurves (rms $\sim$0.01 mag). However, this depth is too deep for a planetary transit although it could be caused by a binary star, in which case it would have already been detected by the Lomb-Scargle periodogram algorithm which was used to identify variable stars in the 47 Tuc field \citep{Weld2004}. On our poor quality lightcurves, this depth is larger than an expected planetary transit.

By incorporating the $\it{W_{i}}$ weighting scheme into the calculation of the $\it{C(P_{mod},\tau_{shift})}$, the number of detections caused by systematic trends was reduced by an order of magnitude. Interestingly, it was found that this weighting scheme only reduces the false detection rates for lightcurves of poor photometric quality. The photometric uncertainty of poor quality lightcurves is strongly dominated by photon noise, which the photometry code can accurately estimate. The point-to-point variations in the fractional photometric uncertainty of bright stars, however, is dominated by systematic effects (such as PSF fitting), which are not accurately reflected in the errorbars returned by most photometry codes. The weighting scheme was used, therefore, only for stars in the lower quality bin.

\section{Transit Recoverability and False Detection Rates}
In order to fully understand the ability of the code to detect transit-like events, extensive Monte Carlo simulations involving modeled transits must be carried out. The limits to which the code can detect transits is a vital part of determining the expected number of planets that potentially could be harvested in any particular dataset, and vital information in placing robust statistical significance on results. With this recoverability knowledge, the false detection rates in the actual dataset can be estimated, which give an indication of the number of candidate detections that will need to be visually examined. For speed of subsequent examination, this false detection rate should be minimized as much as possible.

We stress that using an actual dataset lightcurve instead of a simulated one to carry out the Monte Carlo transit recoverability tests is required. This produces an accurate estimate of the sensitivity of a given algorithm, as the exact temporal sampling and photometric accuracy of the data is reflected in the tests. In summary, many thousands of such time-series transits are produced, and added to actual lightcurves typical of the dataset. The code is then run on all these synthetic transits to see how well the algorithm recovers them with the given detection criteria, allowing a determination of the transit recoverability $\it{R_{tran}}$. The detection criteria can be changed slightly and is subsequently optimised for a given dataset.

The Monte Carlo can also be used to estimate the false detection rate. A false detection is defined as a candidate lightcurve which passes all the detection criteria, yet does not contain a model transit. Systematic effects account for the vast majority of these detections, specifically caused by data points at the beginning and end of a night, by cloudy weather and other spurious terrestrial effects. False detections may also be caused by the statistics of the dataset, as there will always be points that trigger a detection due to random statistics in very large datasets. A further factor to be considered is the effect of stellar crowding on photometry quality. Even with difference imaging techniques, bright stellar companions can seriously degrade the ability of the photometric pipeline to fit a reasonable point spread function to the fainter stars resulting in systematic deficiencies in lightcurve photometry that can easily mimic the appearance of a transit.

False detection rates were determined by running the algorithm with fixed detection criteria on the actual 47 Tuc dataset, and seeing how many ``detections'' the code found. These would contain both real and false detections. The ``real'' transit occurrence rate is expected to be 1/1,250 (from \citet{Weld2005}) and by dividing the transit ``occurrence'' in the code by this number, a value can be estimated for the false detection rate. This value should be minimised in these tests so that the number of candidates for inspection is reduced, yet the number of true detections recovered is large.

\subsection{Recoverability and False Detection Rates in the 47 Tuc dataset}
For the 47 Tuc dataset, a large number of modeled transits were produced in order to test the transit recoverability for the two photometric bins. As the expected depths and durations of planetary transits change as the stellar radius changes along the main sequence of the cluster, a number of different models were produced and tested against the code. A database of 1,464 model transits was produced for a range of orbital periods (60 steps covering $P_{mod}=$1.15 $\rightarrow$ 16.15 day, of 0.25d increment), and model transit start times ($\tau_{mod} =$ 24 steps of 0.5d increment) ranging from the middle to the end of the dataset, for each chosen transit depth and duration to be tested. With four depths-durations combinations, the total database of modeled transits tested with the algorithm numbered 5,856.

For stars in the higher photometric quality bin, model transits were superimposed on a randomly-selected lightcurve with 1$\%$ (0.01 mag) photometric scatter. A dataset was made comprising 1464 models with $\it{D_{mod}=}$ 0.01 mag and a duration $\it{d_{mod}}=$ 2.5 hours, and another 1464 with a $\it{D_{mod}}=$ 0.02 mag and the same duration. The code was then applied to these two model transit datasets, employing the same candidate selection criteria as described in the previous Section. The resultant transit recoverability is shown in Fig.\ref{0.02_rec}. The solid line denotes the ability of the algorithm to recover $\it{D_{mod}}=$ 0.02 mag transits, and the dotted line $\it{D_{mod}}=$ 0.01 mag. The mean of these two recoverabilities is plotted as a lighter shading dotted line. On this scale, a recoverability of 0.5 indicates that half the model transits were detected by the code. A recoverability dropoff is seen as period increases, as these transits have less in-transit points to sample, and have a lower probability of displaying more than one visible transit in the data. Fig.\ref{0.02_rec} indicates that the transit recoverability is very good for these high photometric quality lightcurves over the whole of the sampled orbital period range, with the scatter being due to effects caused by the dataset temporal sampling. 

Fig.\ref{0.04_rec} shows the corresponding recoverability rate for the bin of poorer quality lightcurves, for expected transits of depth $\it{P_{mod}}=$ 0.025 mag and 0.03 mag assuming a duration of $\it{d_{mod}}=$ 2 hours. The solid line indicates the recoverability for a $\it{D_{mod}}=$ 0.025 mag transit added to a lightcurve with rms $=$ 0.025 mag, and the dark shaded dotted line indicates the recoverability for a $\it{D_{mod}}=$ 0.03 mag transit added to a lightcurve with 0.035 mag scatter. Again, for both of these simulations the recoverability is good over all sampled orbital periods, ranging from $\sim$80$\%$ at $\it{P_{tran}} =$ 1 day to $\sim$20$\%$ at 16 days. 

Fig.\ref{weightedmeanrec} incorporates both of the mean recoverabilities derived from all of the simulations, weighted by the numbers of stars in each bin as an indication of the total statistical recoverability in the 47 Tuc dataset. This recoverability was subsequently used to calculate the expected numbers of planets, for a given assumption about planet frequency that should be detected in our data as a whole, as presented in \citet{Weld2005}.

In the better quality bin, it was found that, assuming a planet frequency in 47 Tuc similar to that in the Solar Neighbourhood, a real transit detection should have been identified for every 15 (true and false) detections the code passed, using both the $\sigma_{N_{mod}}$ and $\it{N_{p}}$ detection criteria described earlier. That is, given the number of stars in this bin, 3-4 real transits should be present in a set of 60 candidates, a false detection rate of 15-to-1. Similarly, in the poorer quality bin for which the $\it{W_i}$ weighting scheme was incorporated, the false detection rate was 10-to-1. Although this could have been reduced further by tightening the detection criteria, this was found to impinge too unfavourably on the transit recoverability. Given the number of stars in the second bin, the 50 returned ``detections'' should include 4-5 real planetary transits if 47 Tuc were like the Solar Neighbourhood. 

It is important to note that the proceedure used to weed out false positive detections could conceivably include disregarding real transits. It was found by visual inspection of randomly selected model transits that the signature is clearly visible on the data and has an extremely high detection significance compared to the systematic false positives. It is not expected that the transit recoverability would be significantly impinged by this visual inspection method.

The majority of false detections were caused by crowding and are immediately identified by plotting the lightcurve, which displays systematic features occuring at the same times for most false detections. The problem was less severe in the uncrowded outer regions of 47 Tuc, where the false detection rates were 1/3 the values presented above. Fig.\ref{fdhist} shows the period distribution of all lightcurves that pass the detection criteria and as such is the period distribution of the false detections that were common in the dataset. A few periods have a significant excess of detections over the general trend. In particular $\it{P_{mod}}=$ 1.5 days and $\it{P_{mod}}=$ 4.48 days show a large excess of ``detections''. As it is more difficult for systematics to produce a strong detection when the number of in-transit data points is reduced, the numbers of false detections decreases as period increases.

The code is quick and easy to apply. Application to our whole 47 Tuc dataset was accomplished in 16 steps on four 3057MHz 4096Mb RAM, i86pc machines. The code completed the task in 12 hours, and is easily modifiable to run on existing datasets. 

\section{Summary and Implications for the 47 Tuc Transit Search}
We have presented and described a quick, efficient and easy to apply computational method for the detection of planetary transits in large photometric time-series datasets. Using a cross correlation function the code compares each sampled lightcurve with a database of model transits of appropriate transit depth and duration. A detection is implied by a significantly high value of the correlation distribution. Monte Carlo simulations using the actual temporal sampling and photometric characteristics of the data superimposed on appropriately modeled transits, show an excellent weighted mean recoverability rate over the whole of the sampled period range with a relatively low false detection probability. In particular the code achieves a very good recoverability when searching for transits at or just below the photometric noise level of the data.  The code is easily adaptable to run on existing datasets to search for the same photometric signals, and is capable of testing 10,000 stars in 24 hours with a single 3057MHz machine.

Our algorithm was exclusively developed and tested for use to search an ensemble of 21,950 stars in 47 Tucanae for the presence of Hot Jupiter planets with orbital periods in the range of 1$-$16 days. Despite a detailed search, no transits were found \citep{Weld2005}, yielding a null detection with high significance when compared to the frequency of Hot Jupiters in the Solar Neighbourhood.

\section*{Acknowledgments}
DW wishes to thank the following people for their help in making this paper a reality: Matthew Coleman for answering many questions on the joys of FORTRAN debugging, Terry Bridges for being available to answer questions, and the Mount Stromlo Computer Section for making sure the computers were running continuously in difficult times. Finally, DW would like to thank Ron Gilliland for useful remarks while acting as referee.

\clearpage

\plotone{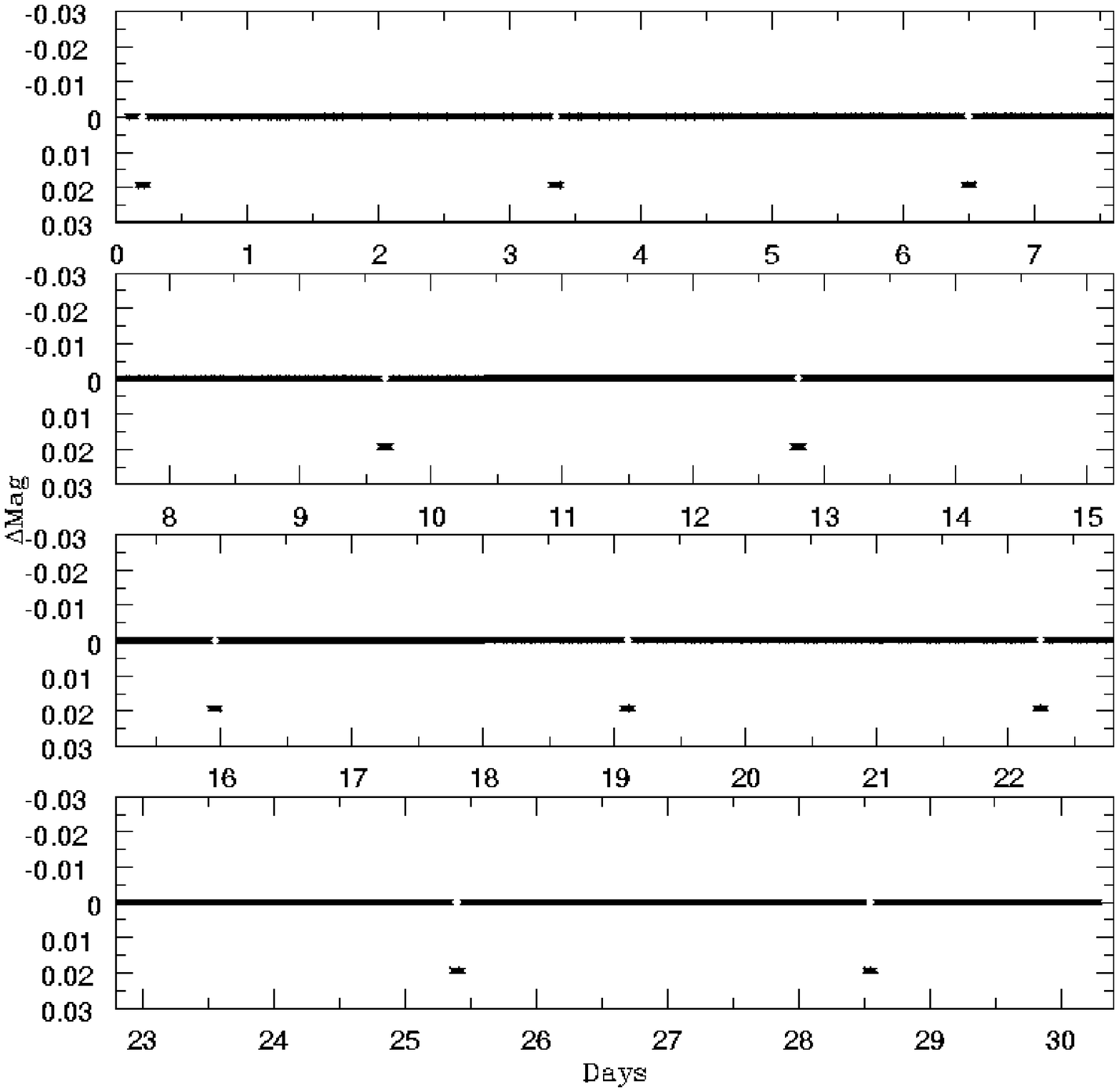}
\figcaption[f1.eps]{An example model transit ($\it{P_{mod}}=$ 3.15 days, $\it{D_{mod}}=$ 0.02 mag) covering the whole of our observing window with no diurnal or weather induced gaps. This shows the ability of a perfect dataset to contain many detectable transits for a typical Hot Jupiter planet.\label{3.15dnogaps}}

\clearpage

\plotone{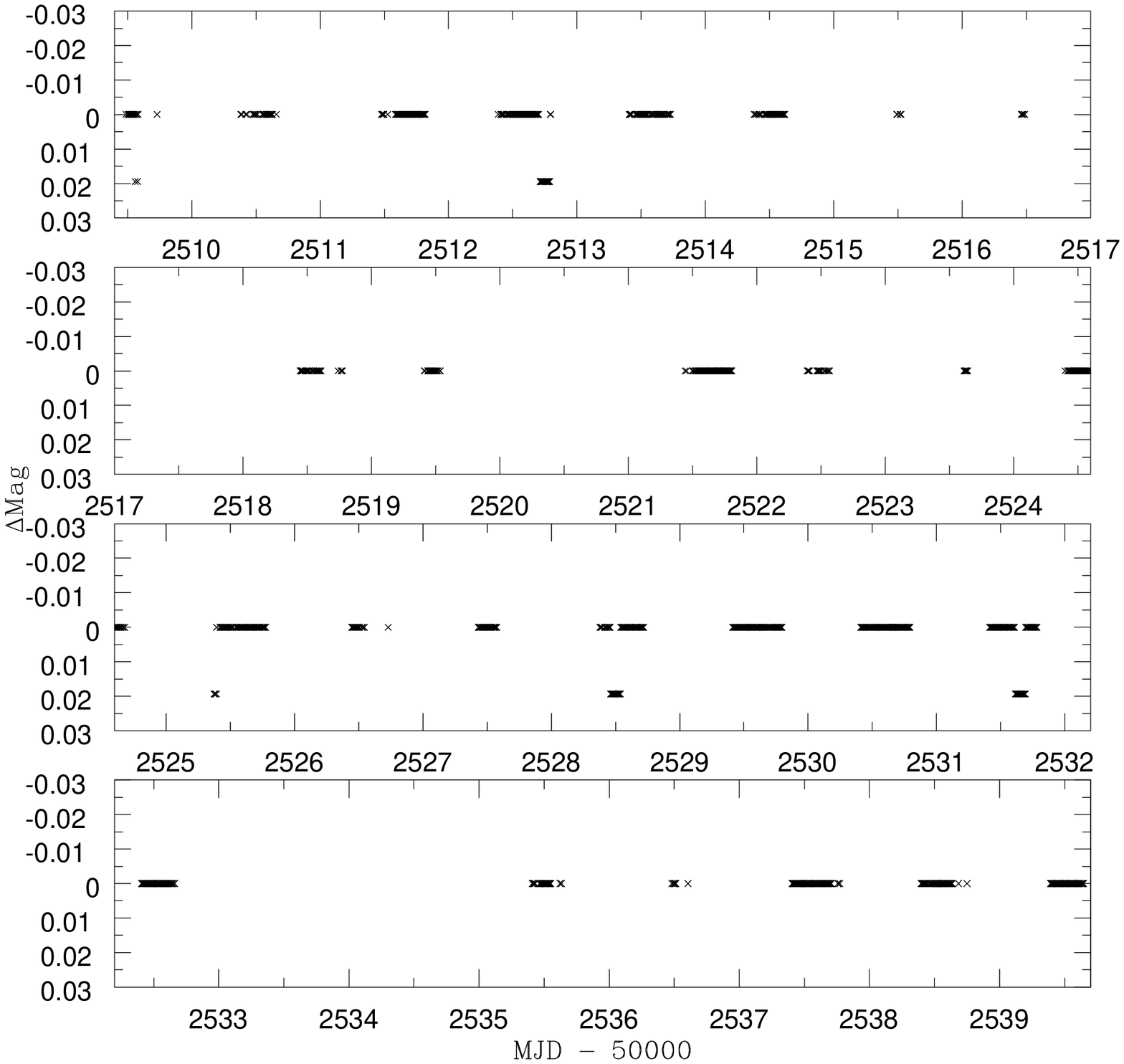}
\figcaption[f2.eps]{The same model transit as seen in Fig.\ref{3.15dnogaps} but with the points shown only for the actual observational times of our 47 Tuc dataset. Gaps caused by weather and daylight eliminate many model transits. Despite this, many detectable transits remain.\label{3.15dgaps}}

\clearpage

\plotone{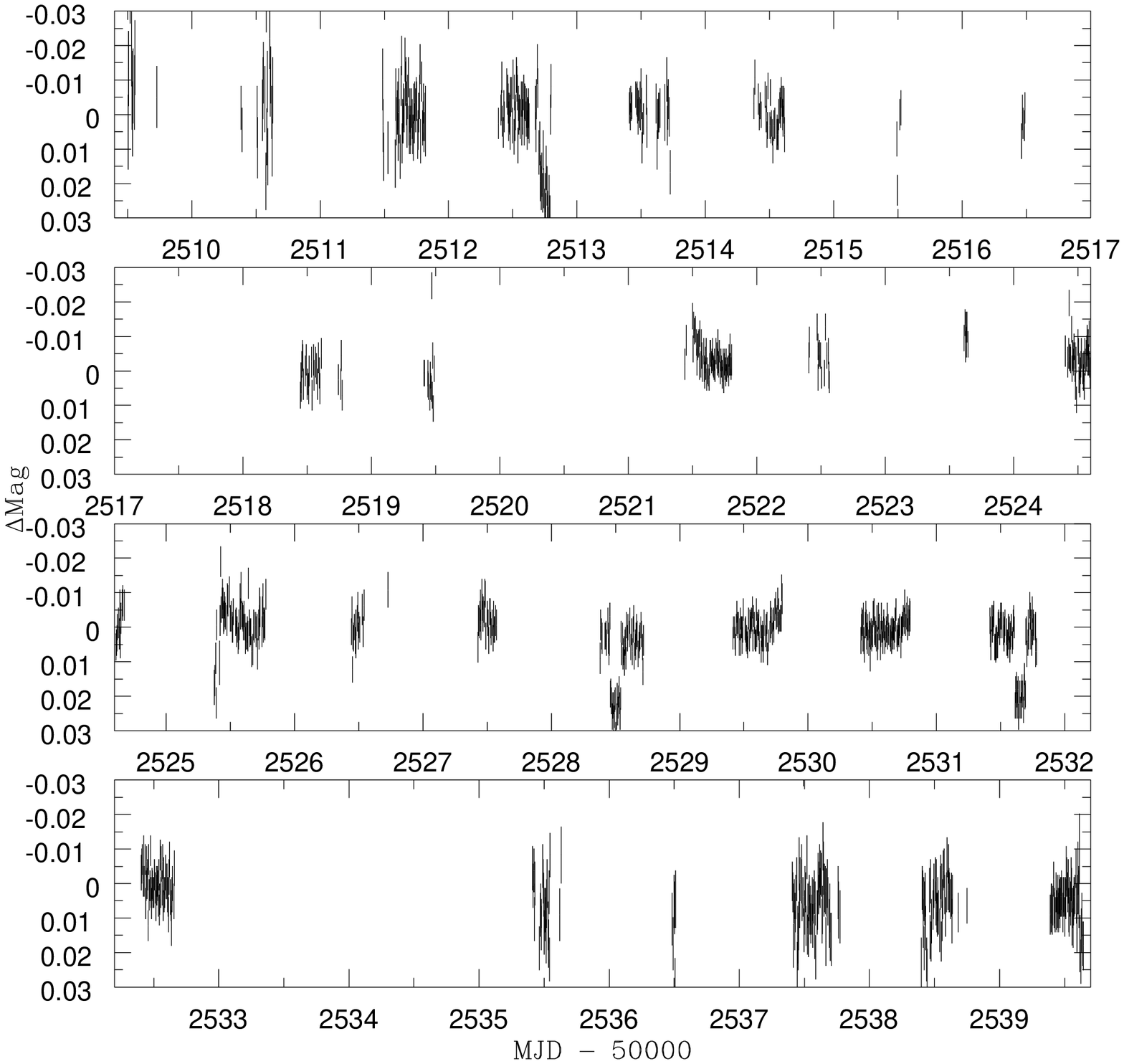}
\figcaption[f3.eps]{A model transit as shown in Fig.\ref{3.15dgaps} sampled at the actual temporal sampling of our data and added to an actual lightcurve from our dataset with rms photometic uncertainty of 0.015 mag. The transit is clearly seen at several places, and the errorbars give an indication of the photometric quality at each observation.\label{3.15ddata}}

\clearpage

\plotone{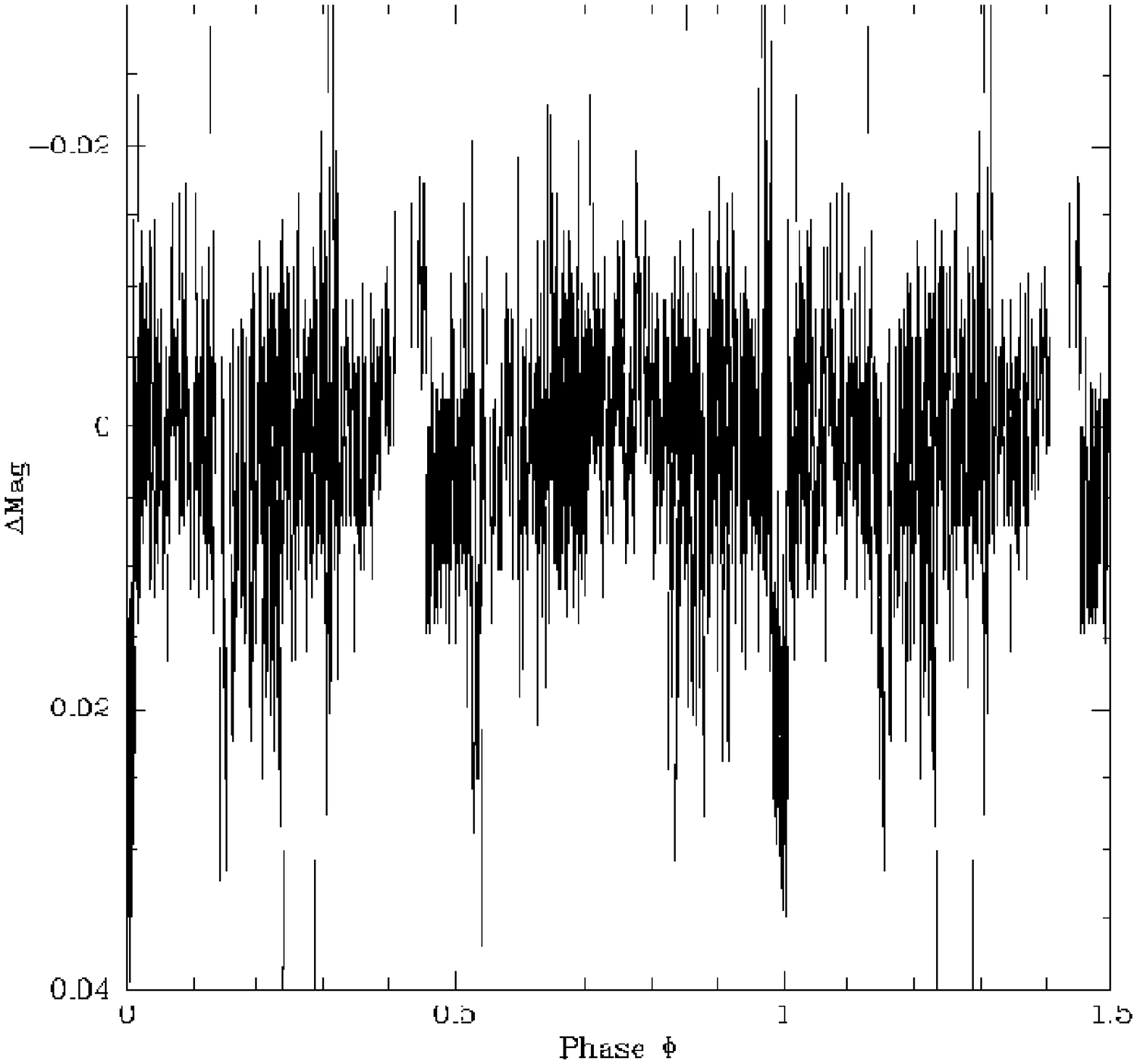}
\figcaption[f4.eps]{The lightcurve shown in Fig.\ref{3.15ddata}, now phase-wrapped to the model period of highest significance ($\it{P_{mod}}=$3.15d) as determined by our detection algorithm. The transit, seen at $\Phi$ $=$ 0 and 1, gives an excellent indication of transit visibility in our 47 Tuc dataset.\label{3.15dpwrap}}

\clearpage

\plotone{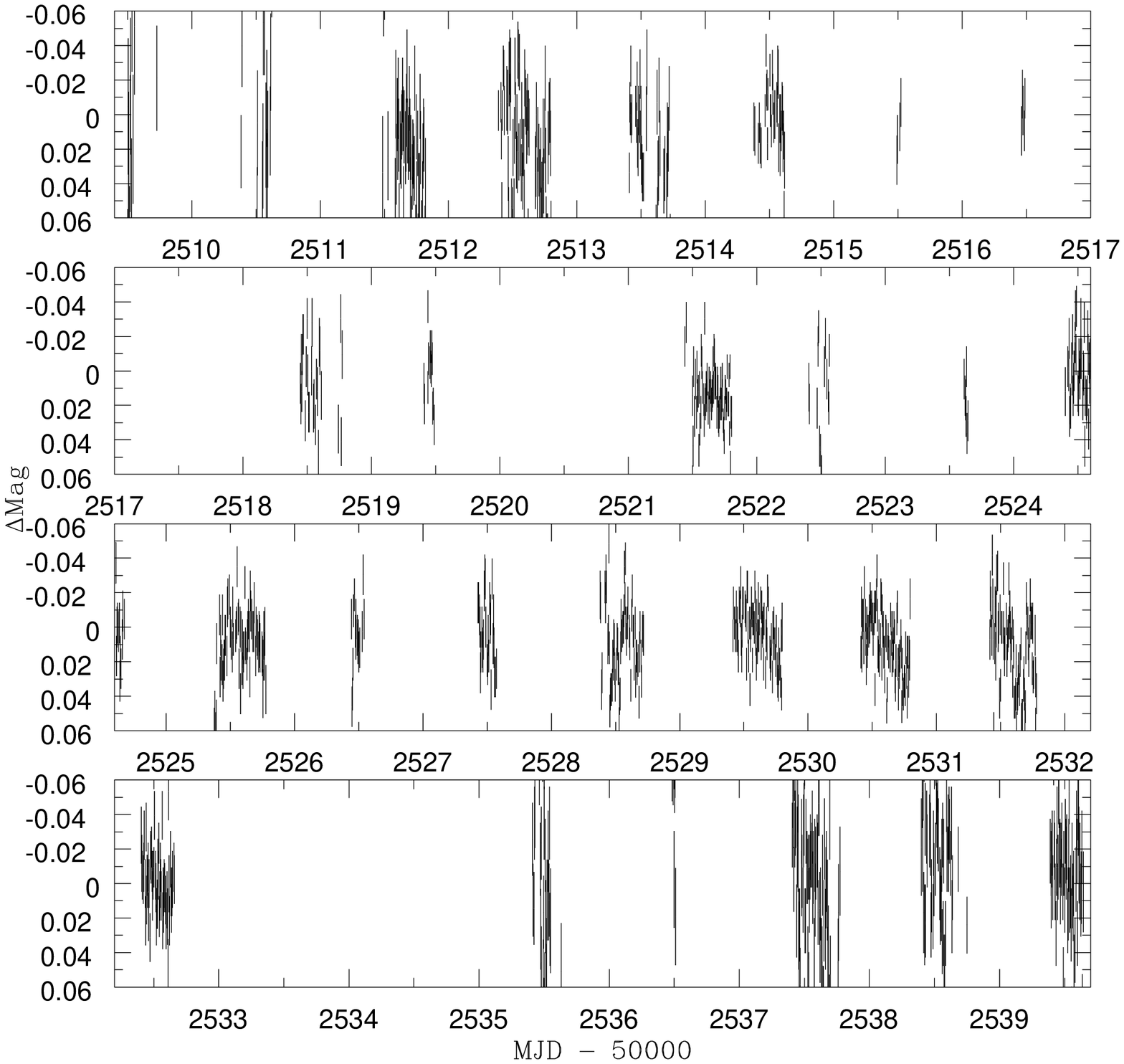}
\figcaption[f5.eps]{The same model $\it{P_{mod}}=$3.15d transit of Fig.\ref{3.15dpwrap} but with a $D_{mod}=$0.03 mag, now superimposed on the lightcurve of a star with rms uncertainty 0.035 mag, typical of the worst quality data searched in the 47 Tuc dataset. Although the transit occurs at the same places as that in Fig.\ref{3.15ddata}, it is much more difficult to see visually. \label{3.15ddata_0.03}}

\clearpage

\plotone{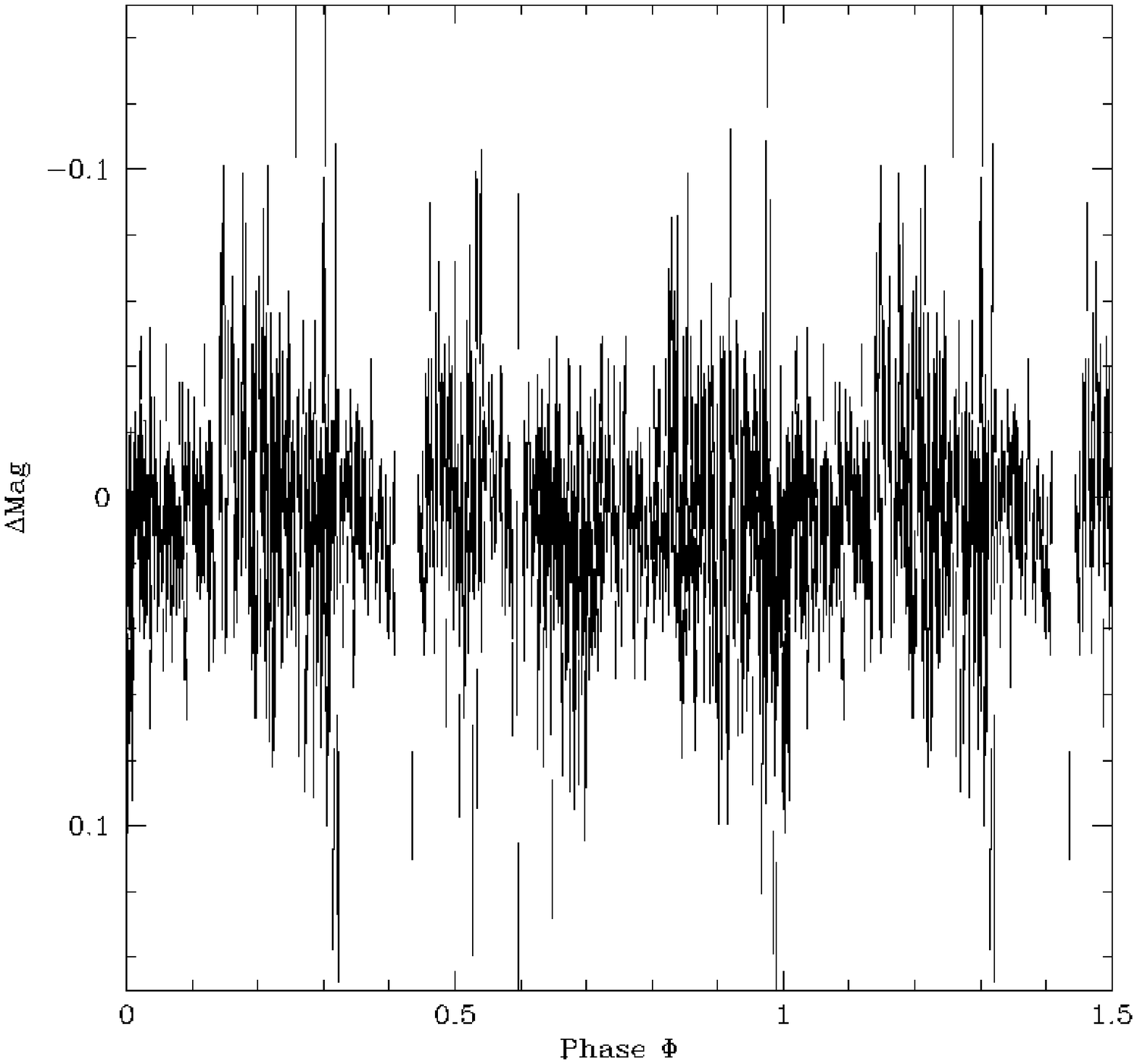}
\figcaption[f6.eps]{The lightcurve of Fig.\ref{3.15ddata_0.03} is phase-wrapped to the model period yielding the highest significance. The transit is barely visible at phase $\Phi$ $=$ 0 and 1. Nevertheless, during the Monte Carlo simulations, the code detected the transit in this lightcurve at the correct period with 9.4$\sigma$ significance.\label{3.15dpwrap_0.03}}

\clearpage

\plotone{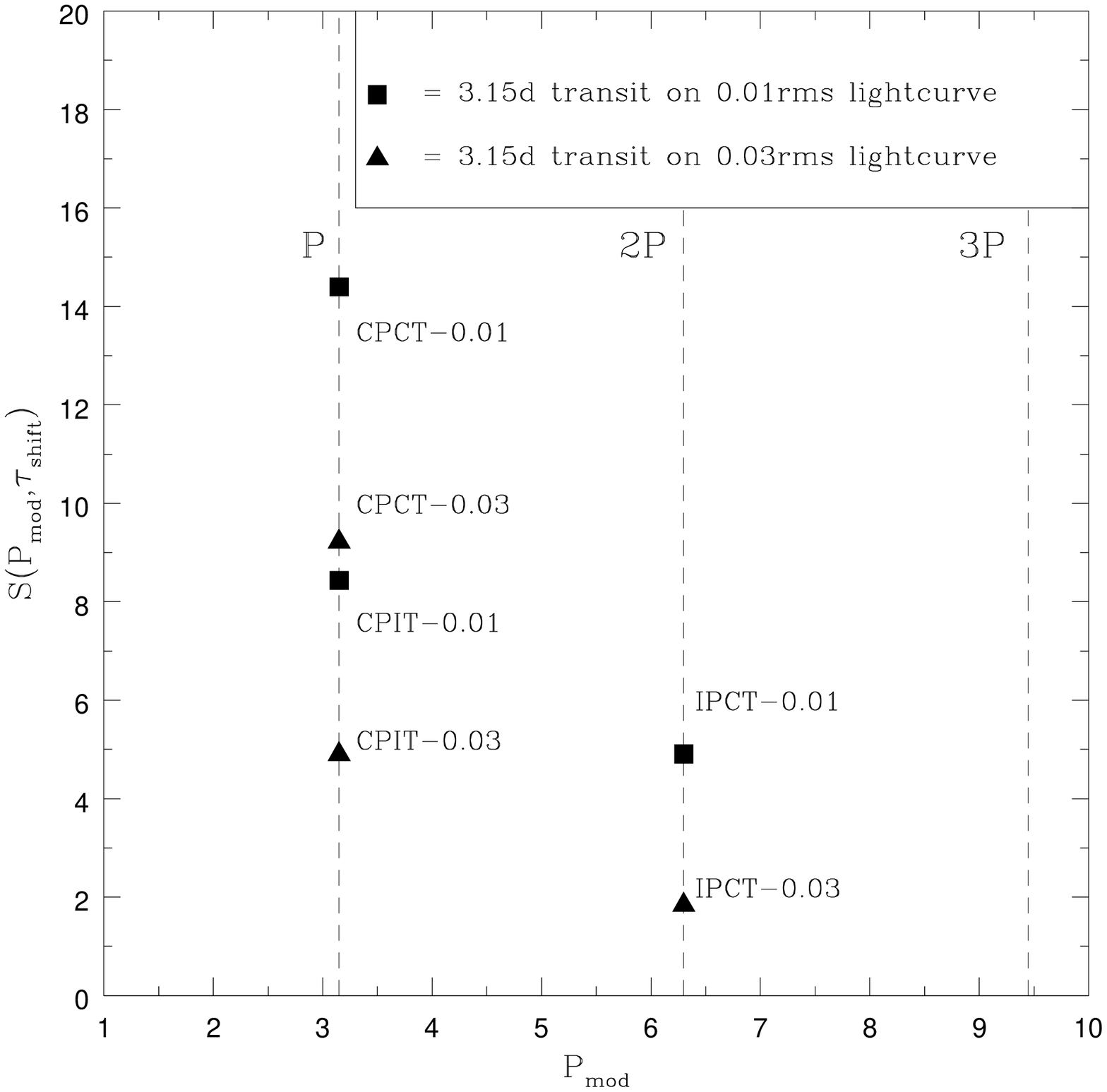}
\figcaption[f7.eps]{The output of the code, using the two lightcurves shown in Figs.\ref{3.15ddata} and \ref{3.15ddata_0.03} as examples. The points show the significance of the detection for both lightcurves at their respective periods, P, and integer multiples of that period.  The statistic $\it{S(P_{mod},\tau_{shift})}$ is shown at the positions for the correct $\it{P_{mod}}$ with correct $\it{\tau_{shift}}$ (CPCT), correct $\it{P_{mod}}$ incorrect $\it{\tau_{shift}}$ (CPIT) and incorrect $\it{P_{mod}}$ correct $\it{\tau_{shift}}$ (CPIT).\label{ccfplot}}

\clearpage

\plotone{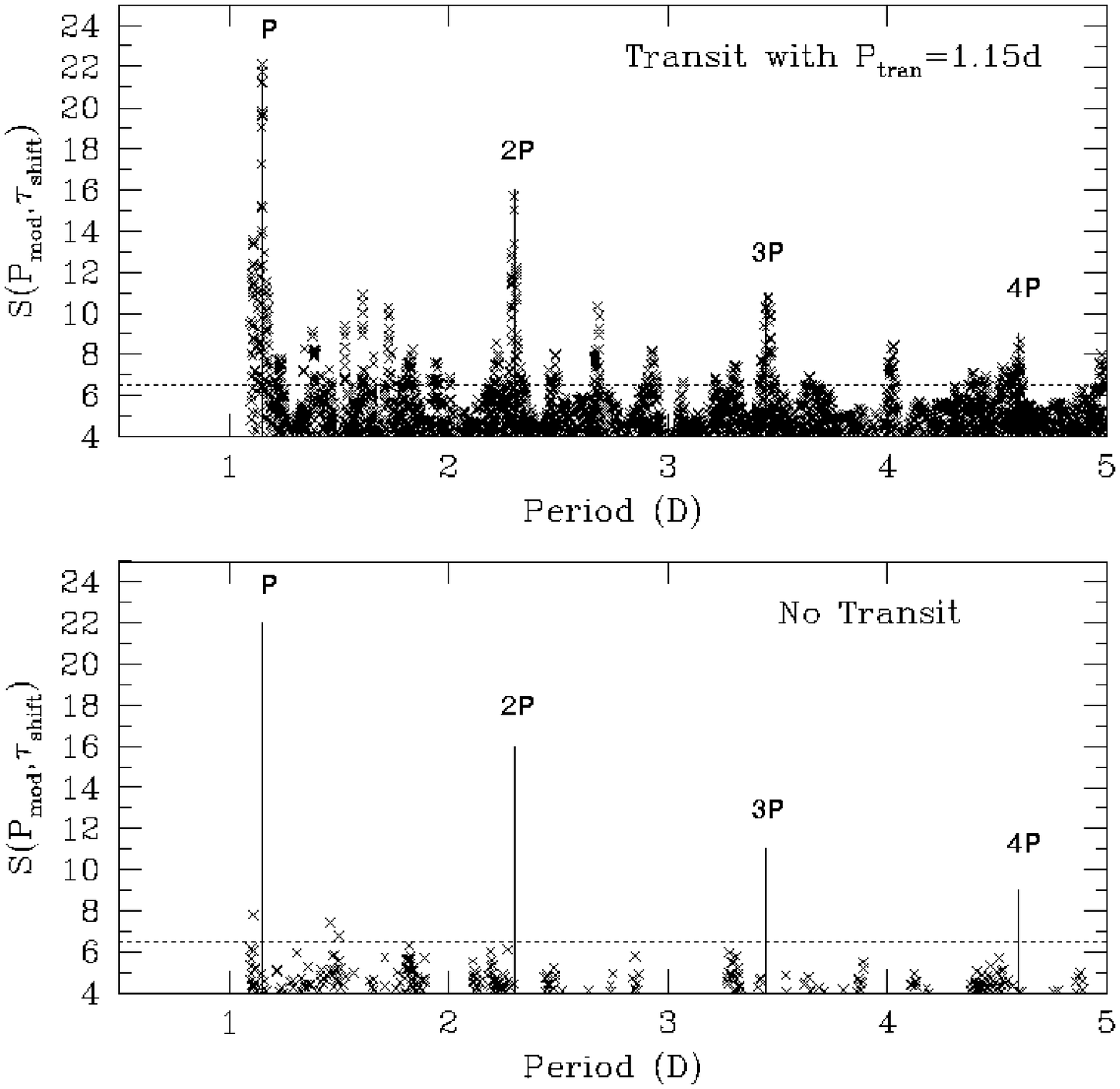}
\figcaption[f8.eps]{The total output of the transit finding code. The top panel shows the output with a $\it{P_{tran} =}$ 1.149 day 2$\%$ (0.02 magnitude) depth transit added to an actual dataset lightcurve with a photometric rms scatter of 0.015 mag. The bottom panel shows the same lightcurve but without the transit. The detection is clear, producing a 22$\sigma$ spike in the output $\it{C(P_{mod},\tau_{shift})}$ values, the actual true period having the highest significance detection, with aliasing apparent at integer multiples of the true period, which are marked. The dotted line indicates the $S_{cr} \ge 6.5$ detection criterion, as described in the text, and as applied to the lightcurves in the bin of higher quality.\label{mfaoutput}}

\clearpage

\plotone{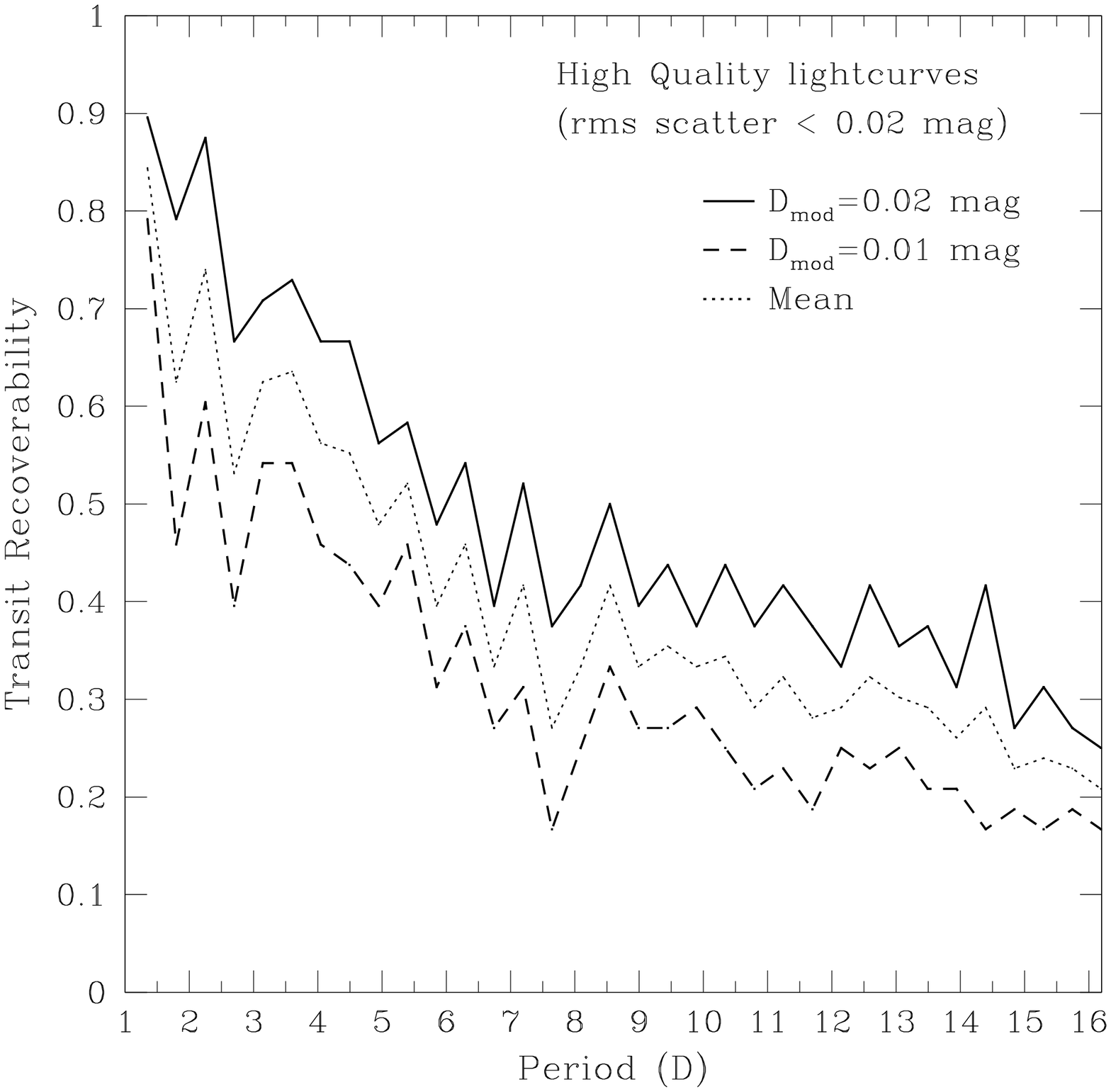}
\figcaption[f9.eps]{The recoverability of $\it{D_{mod}}= $0.01 mag and 0.02 mag transits with duration $\it{d_{mod}}=$ 2.5 hours as a function of orbital period for stars with rms scatter $\le$0.02 mag. The solid line indicates the recoverability of 0.02 mag transits, whereas the dotted line indicates recoverability of 0.01 mag transits. The light dotted line is the mean recoverability of these two simulations.\label{0.02_rec}}

\clearpage

\plotone{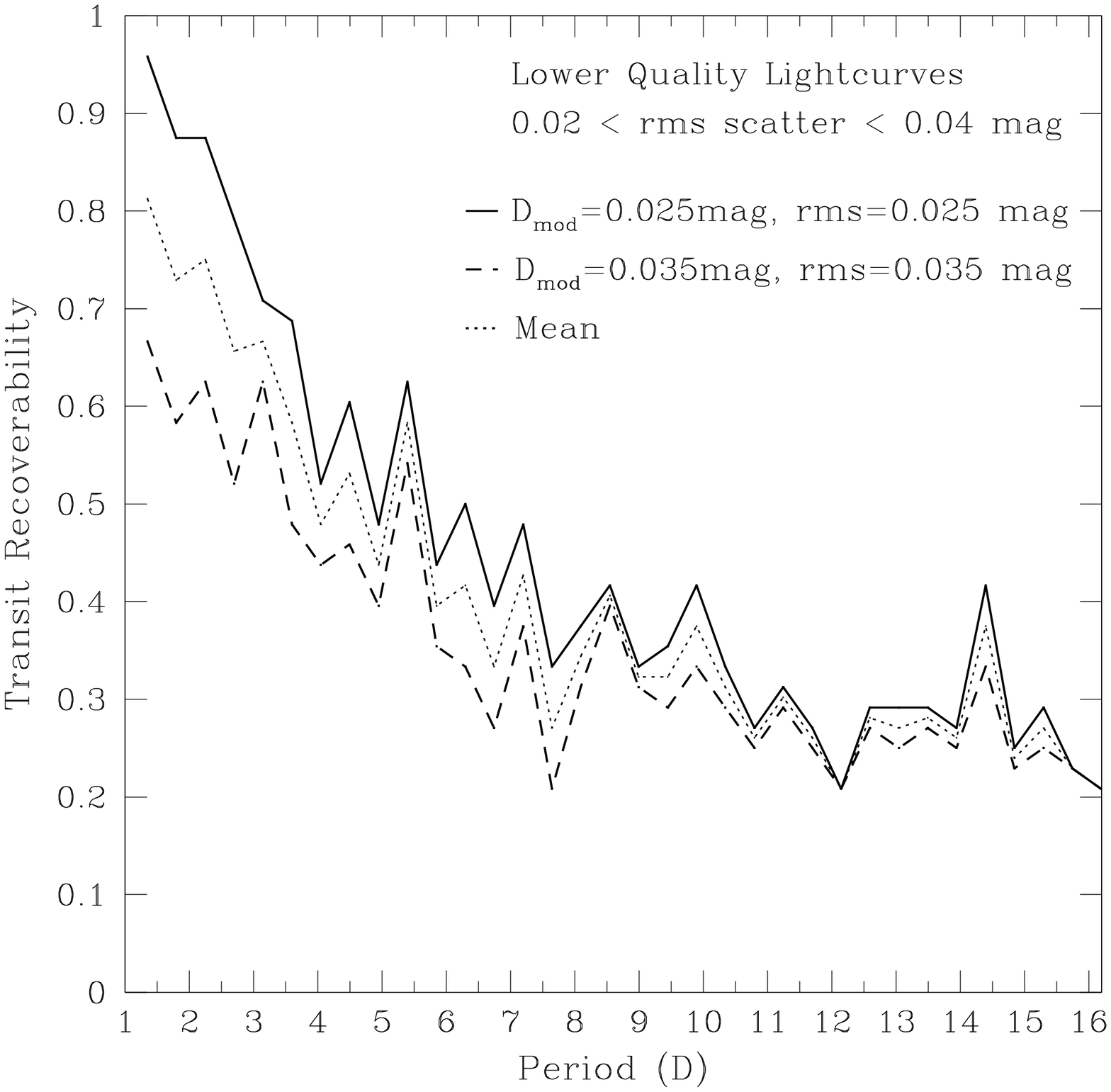}
\figcaption[f10.eps]{The transit recoverability for the second of our dataset bins, stars with rms scatter in the range 0.02-0.04 mag. The grey dotted line indicates the average recoverability for this bin. \label{0.04_rec}}

\clearpage

\plotone{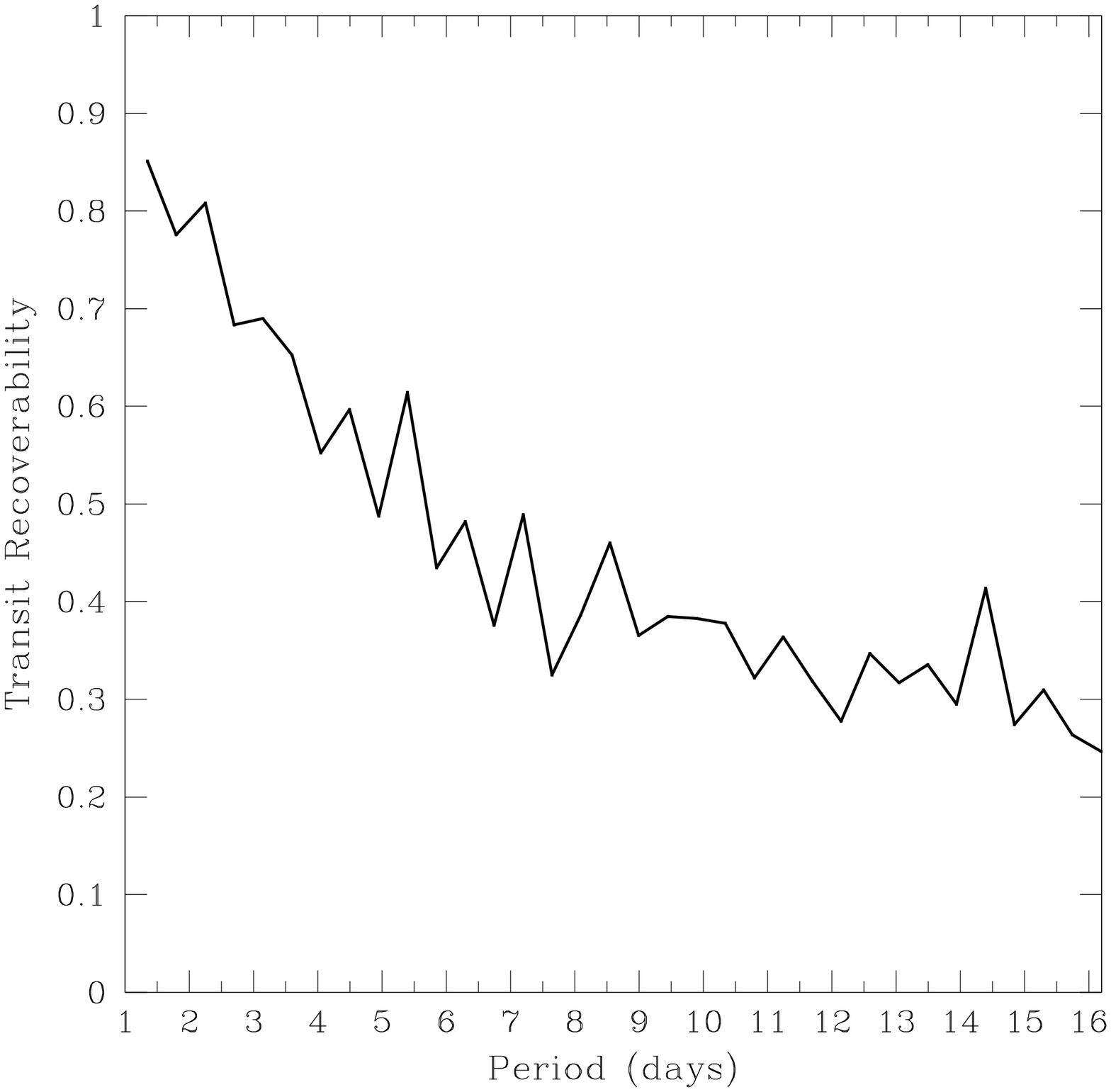}
\figcaption[f11.eps]{The weighted mean transit recoverability as a function of orbital period in our dataset. This is determined via the mean recoverabilities as plotted in Figs.\ref{0.02_rec} and \ref{0.04_rec} and incorporates the number of stars in each of the two magnitude rms bin which shall be sampled in our dataset. This recoverability was used to determine the expected number of planets that should be seen in our 47 Tuc dataset.\label{weightedmeanrec}}

\clearpage

\plotone{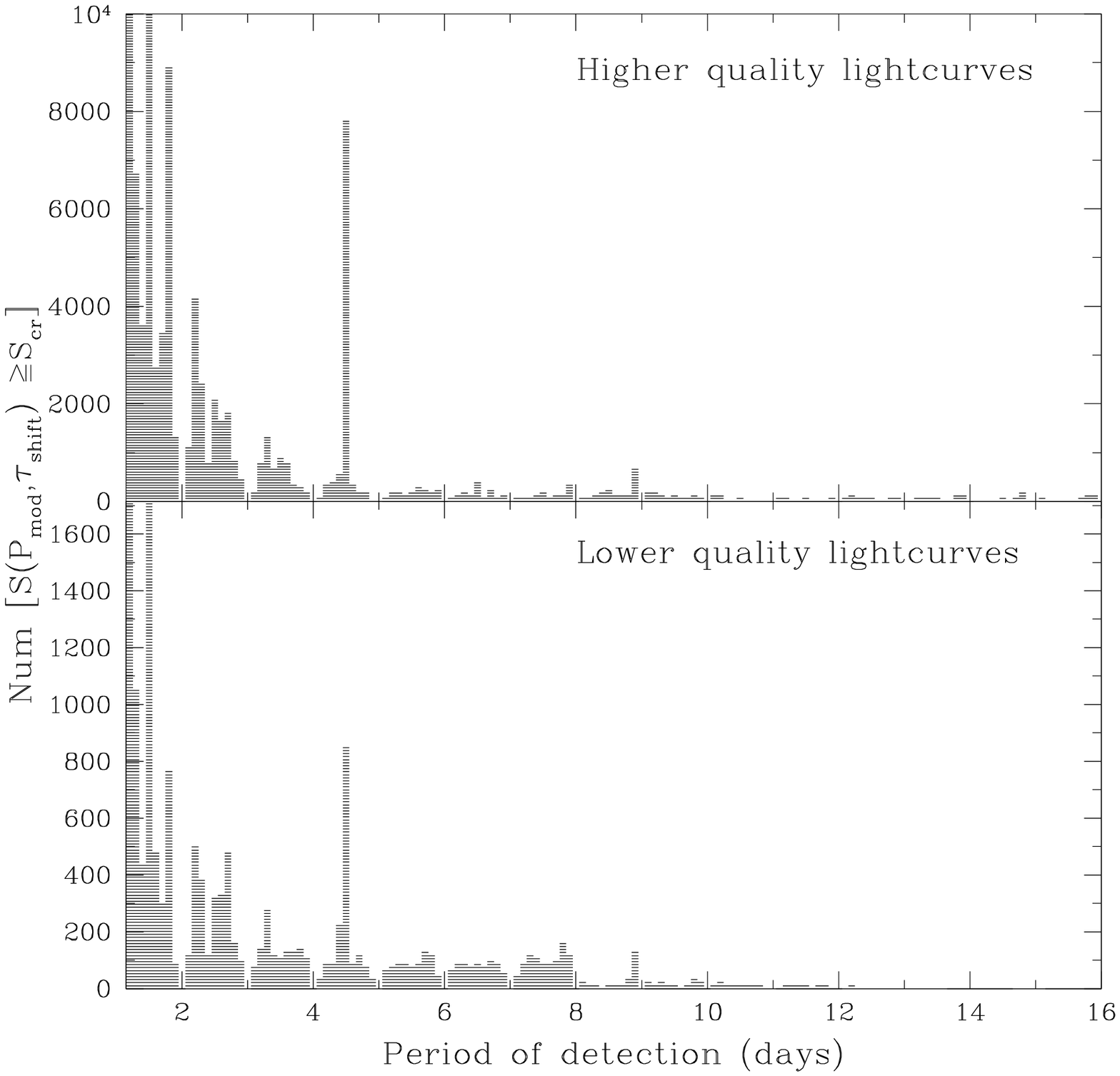}
\figcaption[f12.eps]{The normalised period distribution of all the models, after running the code on the whole dataset, which satisfy the $S(P_{mod},\tau_{shift}) \ge S_{cr}$ detection criterion. Both photometric quality bins are plotted. A few periods (in particular P$=$4.48 days) show many systematic false detections.\label{fdhist}}

\end{document}